\documentclass[aps,prb,twocolumn,superscriptaddress,showpacs]{revtex4}
\makeatletter

\newcommand{\Rmnum}[1]{\expandafter\@slowromancap\romannumeral #1@}
\makeatother

\usepackage{eurosym}
\usepackage{amsfonts}
\usepackage{amssymb}
\usepackage{amsmath}
\usepackage{graphicx}
\usepackage{color}
\begin{document}

\title{Strain-controlled switch between ferromagnetism and antiferromagnetism in 1$T$-CrX$_{2}$ (X = Se, Te) monolayers}

\author{H. Y. Lv}

\affiliation{Key Laboratory of Materials Physics, Institute of Solid State Physics, Chinese Academy of Sciences, Hefei 230031, People's Republic of China}

\author{W. J. Lu}
\email[Corresponding author: ]{wjlu@issp.ac.cn}
\affiliation{Key Laboratory of Materials Physics, Institute of Solid State Physics, Chinese Academy of Sciences, Hefei 230031, People's Republic of China}

\author{D. F. Shao}

\affiliation{Key Laboratory of Materials Physics, Institute of Solid State Physics, Chinese Academy of Sciences, Hefei 230031, People's Republic of China}

\author{Y. Liu}

\affiliation{Key Laboratory of Materials Physics, Institute of Solid State Physics, Chinese Academy of Sciences, Hefei 230031, People's Republic of China}

\author{Y. P. Sun}
\email[Corresponding author: ]{ypsun@issp.ac.cn}
\affiliation{Key Laboratory of Materials Physics, Institute of Solid State Physics, Chinese Academy of Sciences, Hefei 230031, People's Republic of China}
\affiliation{High Magnetic Field Laboratory, Chinese Academy of Sciences, Hefei 230031, People's Republic of China}
\affiliation{Collaborative Innovation Center of Advanced Microstructures, Nanjing University, Nanjing 210093, People's Republic of China}

\makeatletter


\begin{abstract}
We report on the strain-induced switch between ferromagnetic (FM) and antiferromagnetic (AFM) orderings in 1$T$-CrX$_{2}$ (X = Se, Te) monolayers based on the first-principles calculations. The CrSe$_{2}$ and CrTe$_{2}$ monolayers without strains are found to be AFM and FM, respectively. Under the biaxial tensile strain, the CrSe$_{2}$ monolayer tends to be FM when the strain is larger than 2\%. The FM state is further stabilized when the strain is increased. Moreover, the CrSe$_{2}$ monolayer changes to be half-metallic when the tensile strain is larger than 10\%. While for the CrTe$_{2}$ monolayer, the critical strain at which the transition between the FM and AFM states occurs is compressive, of $-1\%$. Relatively small tensile strains of 4\% and 2\%, respectively, can enhance the Curie temperature of CrSe$_{2}$ and CrTe$_{2}$ monolayers above the room temperature. The strain-induced switch between the FM and AFM states in CrSe$_{2}$ (CrTe$_{2}$) monolayer can be understood by the competition between the AFM Cr-Cr direct exchange interaction and FM Cr-Se(Te)-Cr superexchange interaction. The tunable and attractive magnetic and electronic properties controlled by the flexible strain are desirable for the future nanoelectronic applications.

\end{abstract}
\maketitle

\section{INTRODUCTION}

Since the discovery of graphene, which is an atomic layer exfoliated from the graphite, two-dimensional (2D) materials have attracted considerable research interest in recent years, due to their potential applications in the nanoelectronics based on the diverse and attractive properties. Similar to the graphite, most of the layered materials are stacked with their building blocks connected by weak van der Waals interactions. Thus it is possible to obtain few layers or even monolayer from their bulk counterparts.\cite{Science-2011,AdvMatter-2011,NatureCommu-2015} Among various kinds of the layered materials, transition metal dichalcogenides (TMDs) stand out because some of them exhibit even much more exciting properties compared to the star material graphene. For instance, MoS$_{2}$ monolayer was found a semiconductor with sizeable direct band gap of about 1.8-1.9 eV,\cite{PRL-2010,NanoLett-2010,NanoLett-2014} overcoming the disadvantage of zero band gap in graphene when applied in the field effect transistors (FETs) and optoelectronic devices.

With respect to the spintronic applications, 2D ferromagnets with high Curie temperature are desirable. However, neither graphene nor most of the TMDs monolayers is intrinsically magnetic. Many methods can be used to induce the magnetic properties in the TMDs monolayers, mainly including doping,\cite{PRB-2013-Cheng,PRB-2013-Ashwin,PRB-88-144409,JPCM-WSe2} hydrogenation,\cite{PRB-MoS2,JPCC-VS2,SciRep-VTe2,APL-TaS2} and forming zigzag edges.\cite{JAP-WS2,NanoResLett-WS2} In the nanoelectronic applications, however, we expect the magnetism can be precisely and flexibly controlled. These methods mentioned above are somewhat rough as compared with the practical requirements. Recently, VX$_{2}$,\cite{ACSNano-VS2} NbX2,\cite{ACSNano-NbS2} and MnX$_{2}$\cite{PCCP-MnS2} (X = S, Se) monolayers were theoretically predicted to exhibit ferromagnetic (FM) behavior, broadening the properties of the pristine 2D monolayers.

Very Recently, the metastable compound 1$T$-CrTe$_{2}$ was successfully synthesized in the experiment and it exhibits a FM property with a high Curie temperature of 310 K.\cite{JPCM-CrTe2} While it was reported that the CrSe$_{2}$ compound is antiferromagnetic (AFM) and its magnetic property can be tuned by substituting Cr with V or Ti.\cite{PRB-CrSe2} Then what is the origin of the different magnetic properties of CrSe$_{2}$ and CrTe$_{2}$ compounds? On the other hand, 2D materials are needed when applied in the nanoelectronic devices. Since CrX$_{2}$ (X = Se, Te) compounds exhibit various and appealing magnetic behavior, whether they can retain these properties when they are exfoliated into monolayers and whether the magnetic properties can be further tuned by the external method, such as strain?\cite{ACSNano-VS2,ACSNano-NbS2} These questions deserve a careful investigation.

In this work, we investigate the electronic and magnetic properties of 1$T$-CrX$_{2}$ (X = Se, Te) monolayers by using the first-principles calculations. The results show that similar to their bulk counterparts, the pristine CrSe$_{2}$ and CrTe$_{2}$ monolayers are AFM and FM, respectively. Interestingly, we find that the switch between FM and AFM can be realized in both of the two monolayers by applying relatively small strains and the Curie temperatures can be enhanced above the room temperature. The magnetic and electronic properties of CrSe$_{2}$ and CrTe$_{2}$ monolayers can be easily tuned by the external strain, making them promising candidates in the future spintronic applications.

\begin{figure*}
\includegraphics[width=1.6\columnwidth]{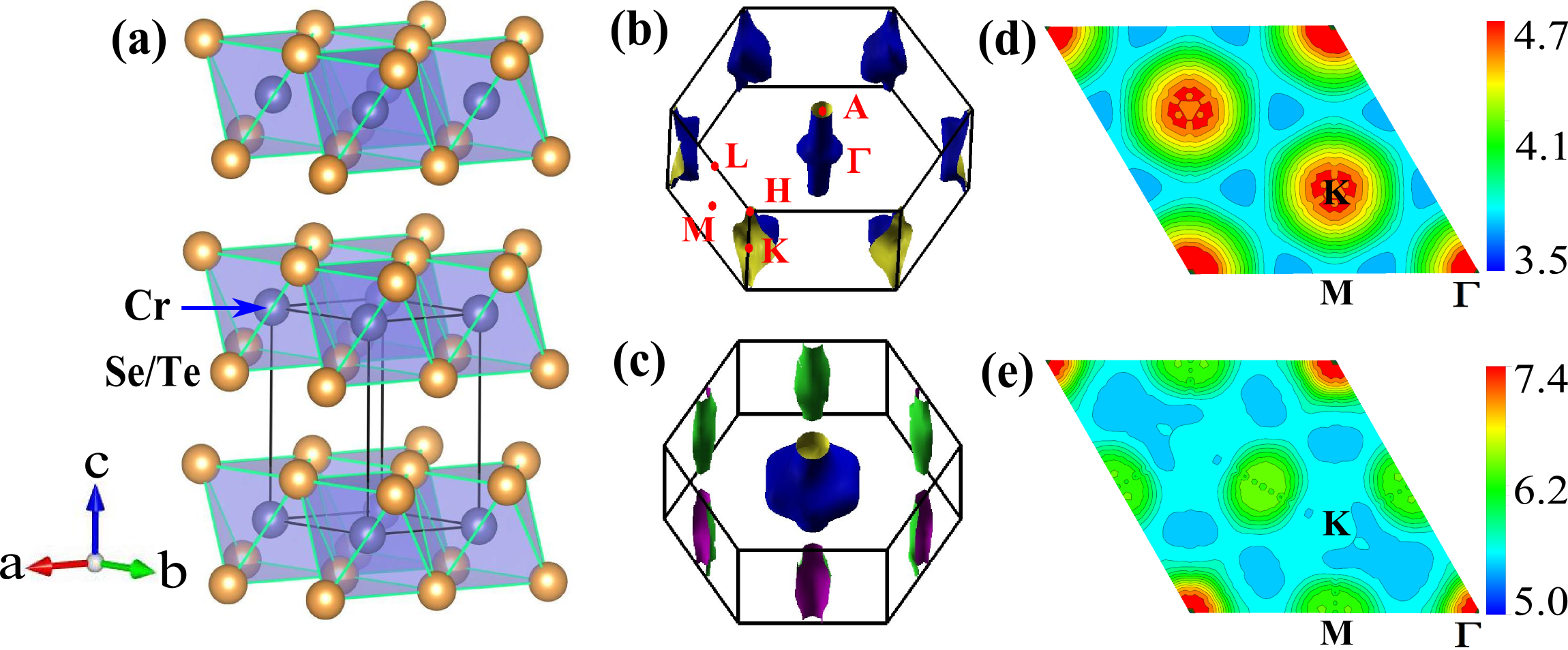}\caption{\label{fig1-energy} (a) Crystal structure of CrSe$_{2}$/CrTe$_{2}$ bulk; Fermi surface without spin polarization for (b) CrSe$_{2}$ and (c) CrTe$_{2}$ bulk; real part of the electron susceptibility $\chi'$ with ${\bf{q}}_z=0$ for (d) CrSe$_{2}$ and (e) CrTe$_{2}$ bulk.}
\end{figure*}

\section{Computational details}

The electronic and magnetic properties were investigated based on the first-principles calculations within the framework of the density functional theory (DFT),\cite{PR-DFT} as implemented in the QUANTUM ESPRESSO code.\cite{QE} The ultrasoft pseudopotentials were used to represent the interaction between electrons and ions. The exchange-correlation potential was in form of the Perdew-Burke-Ernzerhof (PBE) expression\cite{PBE} of the generalized-gradient approximation (GGA). The cutoff energy was set to be 544 eV. For each monolayer, a vacuum region of 15 {\AA} was added so that the interactions between the monolayer and its period image can be neglected. We used $2\times2\times1$ and $3\times3\times1$ supercells for the collinear and noncollinear magnetic calculations, respectively. For the structural relaxations, the Brillouin zones were sampled with $8\times8\times1$ and $4\times4\times1$ Monkhorst-Pack $k$-point meshes for the above two supercells, respectively. For the electronic properties of the collinear calculations, a $10\times10\times1$ Monkhorst-Pack $k$-point mesh was used. As both CrSe$_{2}$ and CrTe$_{2}$ compounds have metastable structures, the lattice parameters were kept fixed at the experimental values when doing the calculations for the bulk.\cite{PRB-CrSe2,JPCM-CrTe2} For the monolayers, when no strain is applied, the in-plane experimental lattice constants of the corresponding bulks were used. The value of the in-plane biaxial strain is defined as $\varepsilon=(a-a_{0})/a_{0}\times100\%$, where $a$ and $a_{0}$ are the in-plane lattice constants of the strained and unstrained monolayers, respectively. At a particular strain condition, the atomic positions were fully relaxed until the force acts on each atom was less than $2.6\times10^{-4} {\mbox{eV}}/{\mbox{\AA}}$.

\begin{figure*}
\includegraphics[width=1.6\columnwidth]{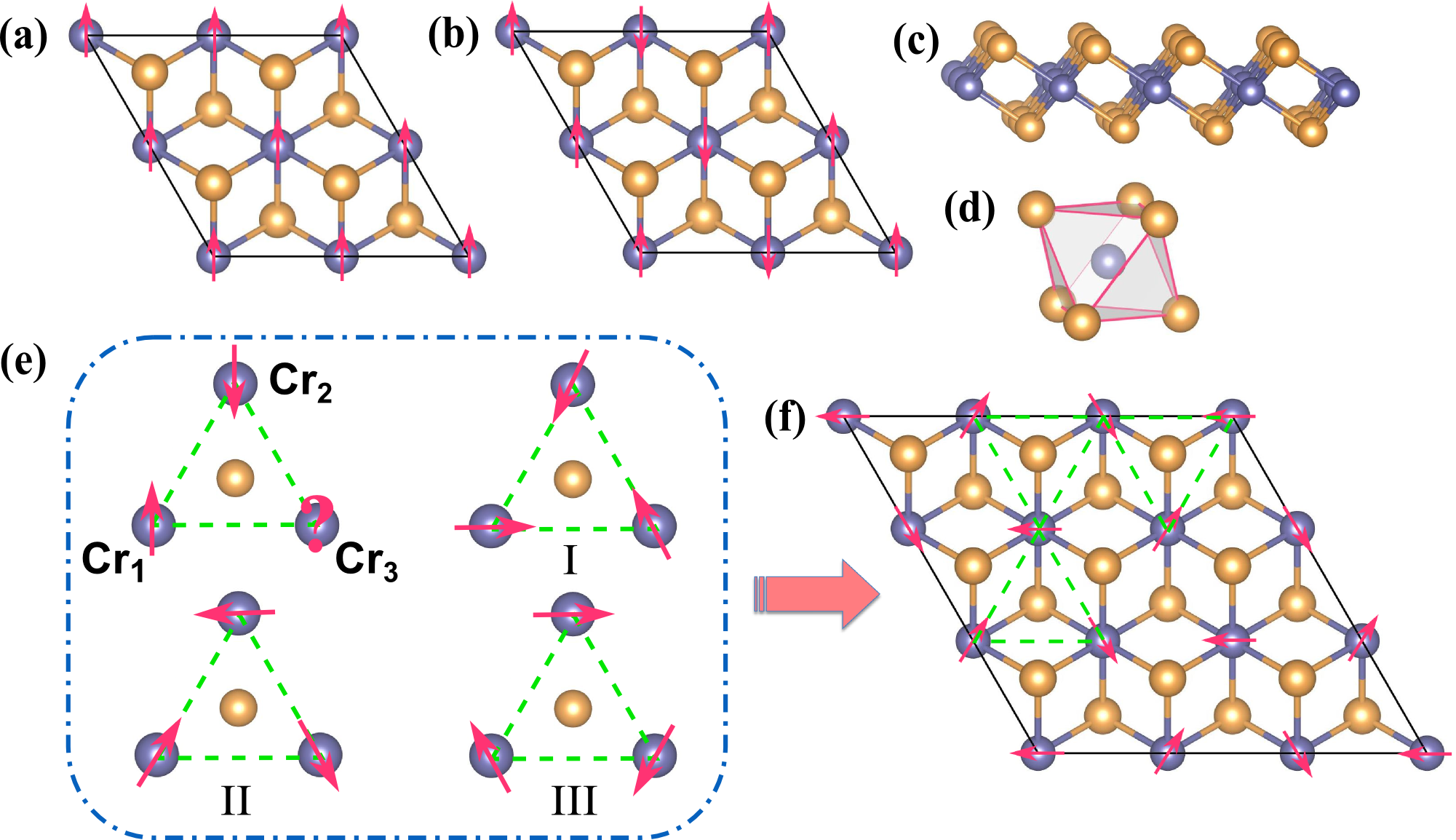}\caption{\label{fig2-structure}Top view of collinear (a) ferromagnetic and (b) antiferromagnetic configurations; (c) side view of CrSe$_{2}$/CrTe$_{2}$ monolayers; (d) the octahedron formed by nearest Se/Te atoms surrounding the Cr atom; (e) (upper left) spin frustration arising from the triangular geometry of the nearest three Cr atoms when their magnetic moments are antiferromagnetically aligned with each other; (the rest) the three solutions of the spin frustrations, with the spin vectors of the three Cr atoms at $120^\circ$ to each other, labeled as ``\Rmnum{1}", ``\Rmnum{2}" and ``\Rmnum{3}", respectively; (f) noncollinear antiferromagnetic configuration of CrSe$_{2}$/CrTe$_{2}$ monolayers constructed by the three solutions in (e). Blue and orange balls represent Cr and Se/Te atoms, respectively.}
\end{figure*}

\section{RESULTS AND DISCUSSION}

\subsection{CrSe$_{2}$ and CrTe$_{2}$ bulks}

The 1$T$-CrSe$_{2}$ and CrTe$_{2}$ compounds were found to have AFM and FM ground states, respectively.\cite{PRB-CrSe2,JPCM-CrTe2} They crystallize in the space group of $P\bar{3}m1$ (No. 164), with the layered structures stacked by the Se-Cr-Se and Te-Cr-Te sandwich layers along the $c$ axis, respectively, as shown in Fig. 1(a). Within the sandwich layer, the Cr atoms are octahedrally coordinated and covalently bonded by the six nearest-neighboring Se/Te atoms, forming a hexagonal layer sandwiched between the two layers of Se/Te atoms. While in between the layers, the interactions are mainly of the van der Waals type.

To reveal the origin of the different magnetic orderings in CrSe$_{2}$ and CrTe$_{2}$ compounds, we first calculate their Fermi surfaces (FS) without spin polarizations, which are shown in Figs. 1(b) and (c), respectively. For the CrSe$_{2}$ bulk, the FS is composed of two hole-like bellied cylinders, one centered along the $\Gamma$-$A$ line and the other along the $K$-$H$ line. The two bellied cylinders indicate that the FS may have a nesting feature, and the nesting vector is along the $\Gamma$-$K$ direction. The FS nesting may be the origin of the magnetic ground state, which will be discussed later. For the CrTe$_{2}$ bulk, however, one hole-like bellied cylinder centered along the $\Gamma$-$A$ line and one electron-like cylinder centered along the $M$-$L$ line form the FS, in good agreement with the previous report.\cite{JPCM-CrTe2} In the $a$-$b$ plane, the FS does not exhibit any nesting property.

The magnetic properties may be controlled by the FS topology,\cite{PRL-Williams} which can be understood by the Fourier transformed exchange coupling constant expressed as\cite{SSP-Kittel,FM-Legvold,PRB-Biasini}
\begin{equation}\label{1}
    J({\bf{q}})\propto\frac{1}{N}\sum_{{\bf{k}}}I({\bf{k}},{\bf{q}})\frac{f(\varepsilon_{\bf{k}})-f(\varepsilon_{\bf{k}+\bf{q}})}{\varepsilon_{\bf{k}}-\varepsilon_{\bf{k}+\bf{q}}}\approx I\chi'(\bf{q}),
\end{equation}

where \begin{equation}\label{2}
    \chi'({\bf{q}})=\frac{1}{N}\sum_{{\bf{k}}}\frac{f(\varepsilon_{\bf{k}})-f(\varepsilon_{\bf{k}+\bf{q}})}{\varepsilon_{\bf{k}}-\varepsilon_{\bf{k}+\bf{q}}}.
\end{equation}
Here ${\bf{q}}$ represents the Fourier component of a generic perturbation, $I({\bf{k}}$) is the exchange integral, which is approximated to a constant $I$ in real cases, $f$ is the Fermi-Dirac distribution function, $\varepsilon_{\bf{k}}$ is the eigenvalue of the electron $\bf{k}$, $\chi'$ refers to the real part of the electron susceptibility and $N$ is the total number of the $\bf{k}$ points. The maximum of $J(\bf{q})$ and thus the maximum of $\chi'$ determine a periodic magnetic moment arrangement with a propagation vector {\bf{q}}. We can deduce from Eqs. (1) and (2) that when the propagation vector {\bf{q}} coincides with the FS nesting vector, the maximum of $\chi'$ appears.

From Fig. 1(d) we can see that for CrSe$_{2}$, the maximum value of $\chi'$ appears at $\bf{q}$=$\frac{1}{3}$$\bf{a^*}$+$\frac{1}{3}$$\bf{b^*}$, conforming the existence of the Fermi-surface nesting in the CrSe$_{2}$ bulk. The nesting vector $\bf{q}$ indicates that there exists a magnetic configuration with a 3$\times$3$\times$1 supercell in the real space, which may correspond to the AFM ground state measured in the experiment.\cite{PRB-CrSe2} However, in the experiment, the details of the spin ordering is not reported. As discussed later, there exists a frustration in the spins of the Cr atoms in this system. A noncollinear AFM configuration with the spin vectors of the nearest-neighboring Cr atoms at $120^{\circ}$ to each other [see Fig. 2 (f)] form a 3$\times$3$\times$1 supercell exactly, consistent with the nesting result, so it may be one possibility of the spin arrangement in this system, which needs to be further checked by the experiment. Actually, this triangle arrangement of the spin structure has been found in LiCrS$_{2}$.\cite{JSSC-Laar} For CrTe$_{2}$ compound, however, except for the maximum near the $\Gamma$ point, we cannot see any other local maximum of $\chi'$ in the (${\bf{q}}_x, {\bf{q}}_y$) plane [see Fig. 1(e)], which may be responsible for the uniform arrangement of the spins (FM configuration) observed in the experiment.

\subsection{CrSe$_{2}$ and CrTe$_{2}$ monolayers}

The interesting magnetic properties of CrSe$_{2}$ and CrTe$_{2}$ bulks inspire us to explore how their monolayer counterparts will perform, which will have potential applications in the nanoelectronics. First, we examine the ground states of CrSe$_{2}$ and CrTe$_{2}$ monolayers. The two collinear FM and AFM magnetic configurations considered in this work are shown in Figs. 2(a) and (b), respectively. Figure 2(c) is the side view of the monolayer, with the Cr layer sandwiched between two Se or Te layers. Each Cr atom is in the center of the octahedron formed by the nearest Se/Te atoms, as shown in Fig. 2(d). The nearest Cr atoms form an equilateral triangle [see upper left corner of Fig. 2(e)]. In the collinear AFM configuration, if we assume the spins of the Cr$_{1}$ and Cr$_{2}$ atoms are up and down, respectively, then the (up or down) spin of the Cr$_{3}$ atom will be the same as the spin of the Cr$_{1}$ or Cr$_{2}$ atom. That's to say, the spin of the Cr$_{3}$ atom cannot align simultaneously antiparallel to the spins of Cr$_{1}$ and Cr$_{2}$ atoms, so it is frustrated. The three solutions of this spin frustration are demonstrated in Fig. 2(e), labeled as ``\Rmnum{1}", ``\Rmnum{2}" and ``\Rmnum{3}", respectively. The spin vectors of the three atoms are at $120^{\circ}$ to each other. Combining these three cases together, we construct a noncollinear AFM configuration with a 3$\times$3$\times$1 supercell, as shown in Fig. 2(f).

\begin{figure}
\includegraphics[width=0.8\columnwidth]{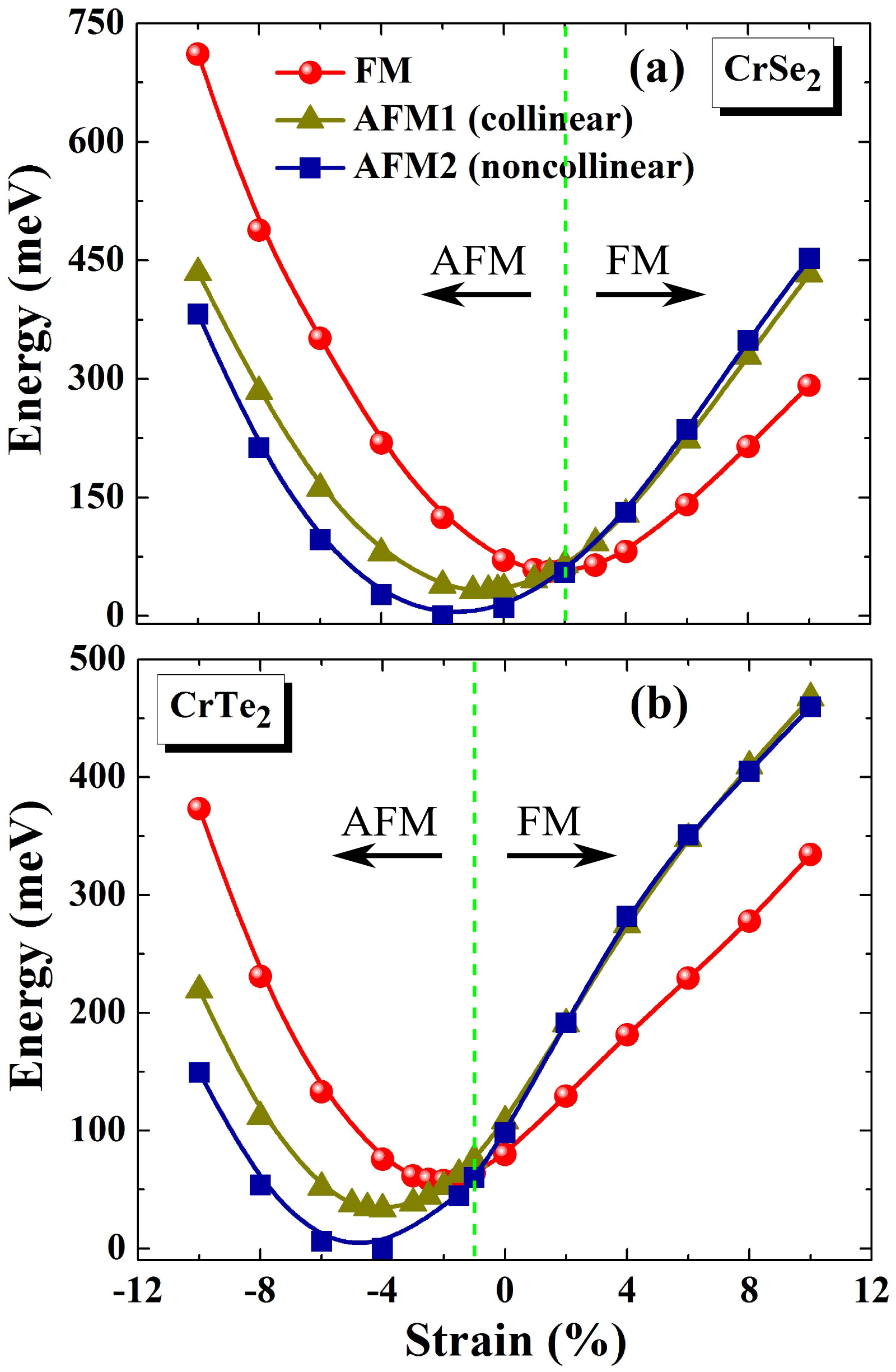}\caption{\label{fig3-energy}Relative total energies of three different magnetic configurations [FM, AFM1 (collinear) and AFM2 (noncollinear)] as a function of the biaxial strain for (a) CrSe$_{2}$ and (b) CrTe$_{2}$ monolayers.}
\end{figure}

The total energy calculations of the three different magnetic configurations [FM, AFM1 (collinear) and AFM2 (noncollinear)] show that the ground states of CrSe$_{2}$ and CrTe$_{2}$ monolayers are AFM and FM, respectively, the same as the results of the corresponding bulk structures. For the CrSe$_{2}$ monolayer, the energy of the noncollinear AFM configuration is relatively lower than that of the collinear one, so the triangular AFM spin arrangement with a 3$\times$3$\times$1 supercell is the ground state of the CrSe$_{2}$ monolayer, which is consistent with the result of the FS nesting in the CrSe$_{2}$ bulk as discussed above. The intrinsic magnetic ground states in the two monolayers are very useful and it is interesting to find ways to further tune the magnetic properties.

Strain is a flexible method to tune the properties of 2D layers.  We then study the strain-dependent magnetic properties of CrSe$_{2}$ and CrTe$_{2}$ monolayers. The in-plane tensile and compressive biaxial strains are applied. The relative total energies of the three different magnetic configurations as a function of the strain are shown in Fig. 3. The energies of the nonmagnetic (NM) systems are much larger than those of the magnetic ones, so they are not displayed here. For CrSe$_{2}$ monolayer [see Fig. 3(a)], the spin ordering tends to be FM when the tensile strain is larger than 2\%. Moreover, the energy difference between the FM and AFM orderings becomes larger and larger as increasing the tensile strain, so the applied tensile strain can stabilize the FM state in this system. On the other hand, the CrSe$_{2}$ monolayer remains AFM when the tensile strain is smaller than 2\% or the applied strain is compressive. So there exists a critical strain of 2\% at which the transition between the FM and AFM states takes place. For the CrTe$_{2}$ monolayer, it remains FM state when the tensile strain is applied. The same as in the case of CrSe$_{2}$ monolayer, the energy difference between the FM and AFM states becomes larger and larger when the tensile strain is increased. The spin ordering changes to be AFM when the absolute value of the compressive strain is larger than 1\%, so the critical strain of CrTe$_{2}$ monolayer is $-$1\%. From the calculated total energies of the three different spin orderings as a function of the strain, we can see that the magnetic states of CrSe$_{2}$ and CrTe$_{2}$ monolayers can be effectively tuned by the external strains. Relatively small strains can induce the transitions between AFM and FM states in the two kinds of materials.

\begin{figure*}
\includegraphics[width=1.6\columnwidth]{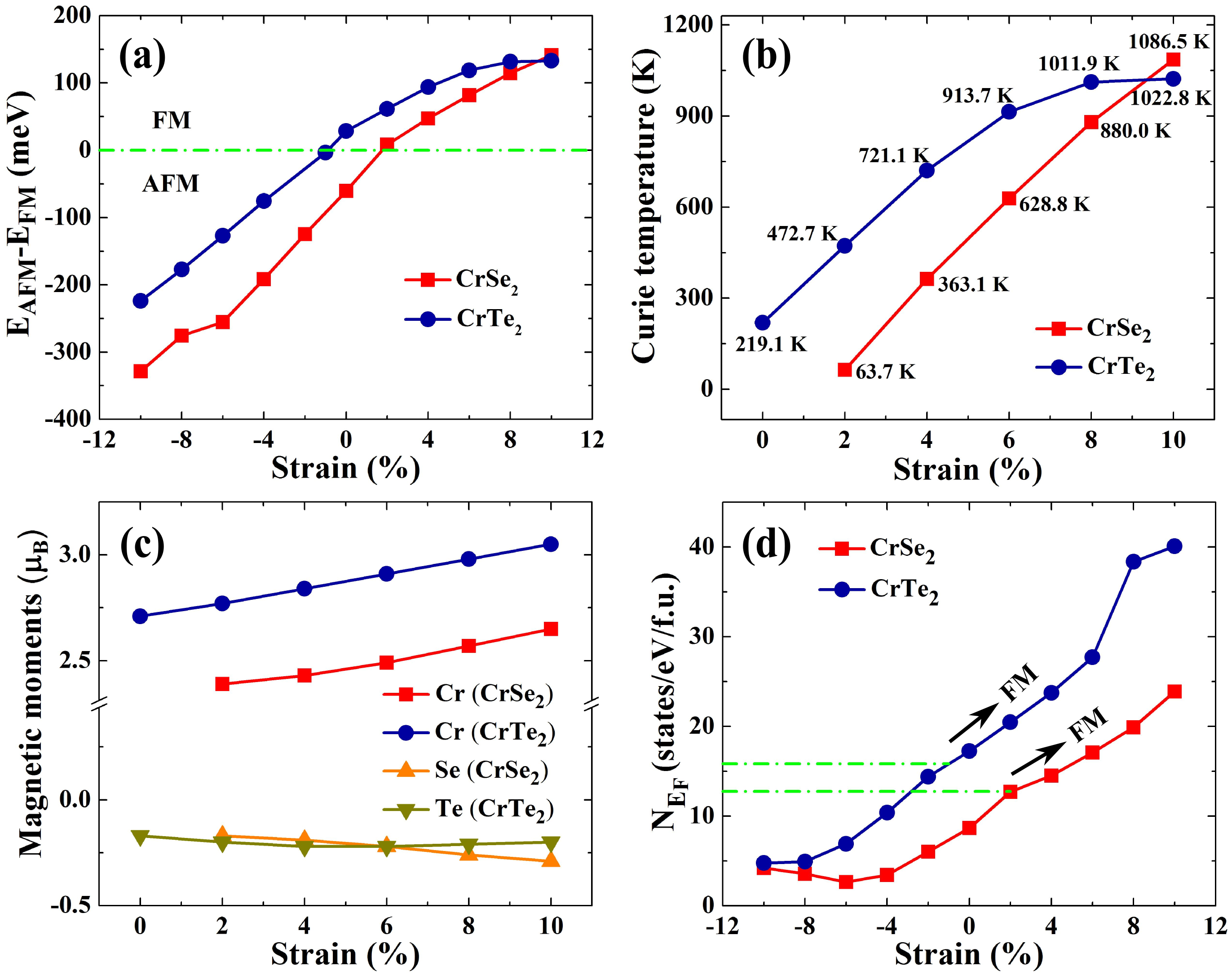}\caption{\label{fig4} Strain-dependence of (a) the energy difference $\Delta$$E$ (= $E_{AFM}-E_{FM}$) between AFM and FM states in one unit cell, (b) the Curie temperature $T_{\mbox{c}}$ in FM states, (c) magnetic moments on Cr and Se/Te atoms in FM states, and (d) the number of the density of states (DOS) at the Fermi energy $N_{E_F}$ in the nonmagnetic (NM) states for CrSe$_{2}$/CrTe$_{2}$ monolayers.}
\end{figure*}

As both CrSe$_{2}$ and CrTe$_{2}$ monolayers can exhibit FM properties, it is interesting to evaluate the Curie temperature $T_{\mbox{c}}$, which is an important design parameter for practical usage. The $T_{\mbox{c}}$ of CrSe$_{2}$/CrTe$_{2}$ monolayers are estimated based on the mean-field theory and Heisenberg model,\cite{PRB-2013-Cheng,PRB-MoS2,Tc-PRB-2004} where $T_{\mbox{c}}$ is expressed as $k_BT_{\mbox{c}}=(2/3){\Delta}E$. Here $k_B$ is the Boltzmann constant and ${\Delta}E$ (= $E_{AFM}-E_{FM}$) is the energy difference between AFM and FM states in one unit cell. The calculated $\Delta$$E$ and $T_{\mbox{c}}$ of CrSe$_{2}$/CrTe$_{2}$ monolayers under different biaxial strains are shown in Figs. 4(a) and (b), respectively. We can see that for both CrSe$_{2}$ and CrTe$_{2}$ monolayers, the $\Delta$$E$ in the FM states and therefore the $T_{\mbox{c}}$ increase as increasing the strain. For CrSe$_{2}$ monolayer, the $T_{\mbox{c}}$ is 63.7 K at the strain of 2\%. When the strain reaches 4\%, the $T_{\mbox{c}}$ increases rapidly to 363.1 K, larger than the room temperature. At the largest strain of 10\%, the $T_{\mbox{c}}$ as high as 1086.5 K can be obtained. For the CrTe$_{2}$ monolayer, as discussed above, it exhibits FM behavior when no strain is applied. The calculated energy difference ${\Delta}E$ is 28.5 meV and $T_{\mbox{c}}$ is 219.1 K, which is smaller than that of the CrTe$_{2}$ bulk (310 K).\cite{JPCM-CrTe2} However, when a small strain of 2\% is applied, the $T_{\mbox{c}}$ can be enhanced above the room temperature, i.e., 472.7 K. At the strain of 10\%, the $T_{\mbox{c}}$ of CrTe$_{2}$ monolayer can reach as high as 1022.8 K. The magnetic moments of Cr ($\mu_{Cr}$) and Se/Te atoms ($\mu_{Se}/\mu_{Te}$) for CrSe$_{2}$/CrTe$_{2}$ monolayers are shown in Fig. 4(c). We can see that for both of the two materials, $\mu_{Cr}$ are much larger than $\mu_{Se}$ or $\mu_{Te}$ and increase monotonically with the increase of the applied strains.

In what follows, we will discuss the origin of the strain-induced magnetic evolution in CrSe$_{2}$/CrTe$_{2}$ monolayers. First, the transition between AFM and FM states can roughly be explained by the Stoner criterion which is expressed as
\begin{equation}\label{1}
   IN_{E_F}>1,
\end{equation}
where $N_{E_F}$ is the number of the density of states (DOS) at the Fermi energy in the NM state. Figure 4(d) shows the calculated $N_{E_F}$ as a function of the strain for CrSe$_{2}$ and CrTe$_{2}$ monolayers. It is clearly seen that the $N_{E_F}$ generally increases as increasing the strain. At the critical strain where the transition between the AFM and FM states occurs, the values of $N_{E_F}$ are 12.7 and 15.8 for CrSe$_{2}$ and CrTe$_{2}$ monolayers, respectively. When the strain is further increased, the $N_{E_F}$ increases as well, so the FM state becomes more and more stable.

\begin{figure}
\includegraphics[width=0.8\columnwidth]{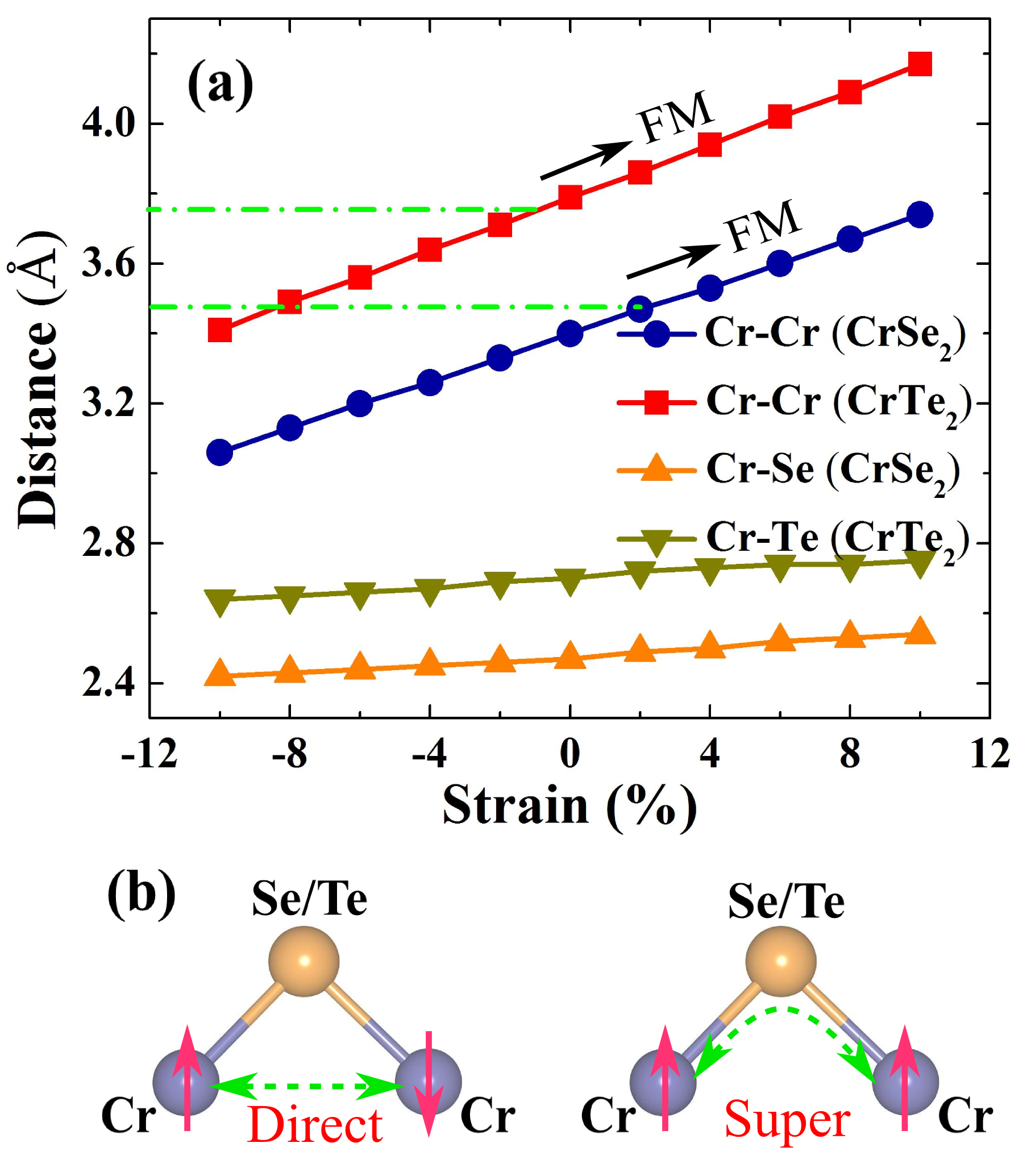}\caption{\label{fig5}(a) Distance between the nearest neighboring Cr atoms $d_{Cr-Cr}$ and bond length between the Cr and Se (Te) atoms $d_{Cr-Se}$ ($d_{Cr-Te}$) as a function of the applied biaxial strain for CrSe$_{2}$ and CrTe$_{2}$ monolayers; (b) Illustrations of the Cr-Cr direct exchange (left panel) and Cr-Se(Te)-Cr superexchange (right panel) interactions in CrSe$_{2}$/CrTe$_{2}$ monolayers.}
\end{figure}

On the other hand, we notice that when the strain is increased, the distance between a Cr atom and its nearest neighboring Cr atom (denoted as $d_{Cr-Cr}$) increases dramatically with the increase of the strain, as shown in Fig. 5(a). For the CrSe$_{2}$ (CrTe$_{2}$) monolayer, $d_{Cr-Cr}$ changes from 3.40 {\AA} (3.79 {\AA}) without strain to 3.74 {\AA} (4.17 {\AA}) at the strain of 10\%, both increased by 10\% when the strain is increased by 10\%. However, the bond lengths between the Cr and Se (Te) atoms [denoted as $d_{Cr-Se}$ ($d_{Cr-Te}$)] increase slowly as increasing the strain. For the CrSe$_{2}$ (CrTe$_{2}$) monolayer, $d_{Cr-Se}$ ($d_{Cr-Te}$) changes from 2.47 {\AA} (2.70 {\AA}) without strain to 2.54 {\AA} (2.75 {\AA}) at the strain of 10\%, increased only by 2.83\% (1.85\%) when the strain is increased by 10\%. This property can be ascribed to the sandwiched structure of CrSe$_{2}$/CrTe$_{2}$ monolayers. When the system is stretched in-plane, $d_{Cr-Cr}$ increases accordingly since all the Cr atoms are in the same layer, while the outside Se/Te atom layers move closer to the Cr atom layer, keeping the $d_{Cr-Se}$ ($d_{Cr-Te}$) slightly changed. For the graphene, since all the carbon atoms are in the same layer, the bond length increases monotonically by 10\% when the tensile strain of 10\% is applied. The relatively small change in the bond lengths in CrSe$_{2}$/CrTe$_{2}$ monolayers indicates they can withstand relatively larger tensile strains before the structural disruption and thus provide a possibility to well control the magnetic properties by a wide range of the external strain.

Based upon the different behaviors of $d_{Cr-Cr}$ and $d_{Cr-Se}$ ($d_{Cr-Te}$) upon the applied strain, the mechanism of the transition between FM and AFM states and the strain-tunable stability of the FM states can be understood by the competition between two different exchange interations. First, since the $d_{Cr-Cr}$ is not very large, the direct exchange interaction between the two nearest-neighboring Cr atoms (denoted as $J_D$) cannot be neglected, which leads to the AFM arrangement of the two Cr atoms [see the left panel of Fig. 5(b)]. Second, as shown in the right panel of Fig. 5(b), the two nearest-neighboring Cr atoms are connected by a Se/Te atom, thus the superexchange interation mediated by the middle Se/Te atom exists. Since the Cr-Se(Te)-Cr bond angle is close to $90^\circ$ and the number of the 3$d$ electrons of the Cr atom is 5, according to the Goodenough-Kanamori-Anderson (GKA) rules,\cite{Goodenough-1955,Kanamori-1960,Anderson-1959} the Cr-Se(Te)-Cr superexchange interaction (denoted as $J_S$) will induce the FM arrangement of the two nearest-neighboring Cr atoms. Therefore, the magnetic states of CrSe$_{2}$/CrTe$_{2}$ monolayers are determined by $J_D+J_S$, where $J_D$ and $J_S$ are negative and positive, respectively. For the CrSe$_{2}$ monolayer without strain, $|J_D|$ is relatively larger than $J_S$, so $J_D+J_S<0$ and the system exhibits AFM property. When the tensile strain is applied, as discussed above, the $d_{Cr-Cr}$ increases much faster than $d_{Cr-Se}$, so $|J_D|$ decreases much faster than $J_S$. At the critical strain of 2\%, $J_S$ is equal to $|J_D|$ and $J_D+J_S=0$. When the strain is larger than 2\%, $J_S$ prevails in determining the magnetic property and $J_D+J_S>0$, so the CrSe$_{2}$ monolayer becomes FM. Moreover, since $|J_D|$ decreases much faster than $J_S$ as increasing the strain, $J_D+J_S$ increases as a function of the strain when the strain is further increased. As a result, the energy difference between FM and AFM states becomes more and more large and the FM state is stabilized by the applied tensile strain. The case for the CrTe$_{2}$ monolayer is very similar, except that the critical strain is compressive, of $-1\%$. For the CrTe$_{2}$ monolayer without strain, the $|J_D|$ is relatively smaller than $J_S$, so $J_D+J_S>0$ and the system is FM. From the above discussion, we can see that the $d_{Cr-Cr}$ plays an important role in determining the magnetic ground states of CrSe$_{2}$/CrTe$_{2}$ monolayers, which has also been found in other Cr-based systems.\cite{JPCM-1996,JPCM-1997,PLA-Chen}

\begin{figure}
\includegraphics[width=1.0\columnwidth]{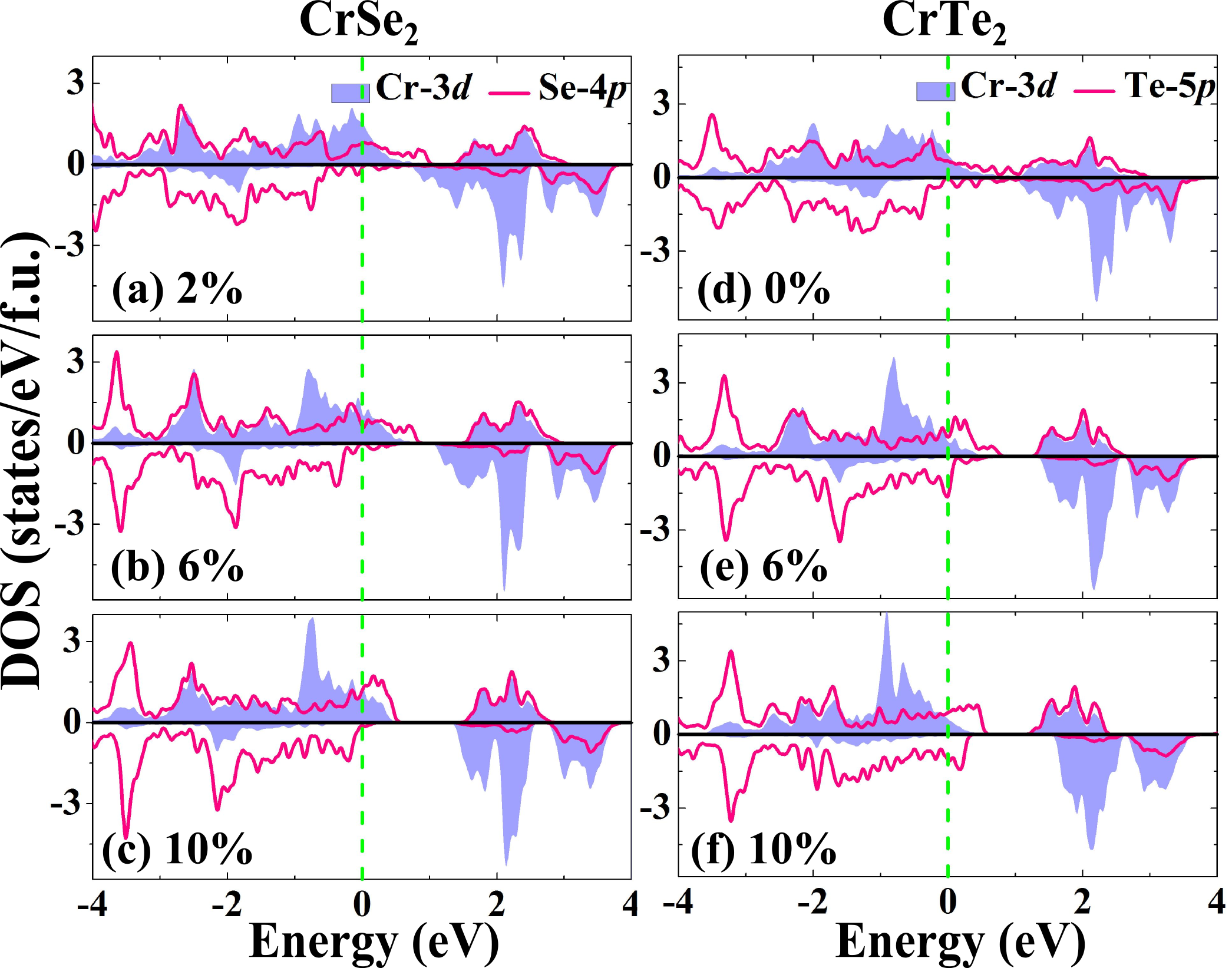}\caption{\label{fig6-DOS}Partial density of states (PDOS) of Cr 3$d$ and Se 4$p$ orbitals for the CrSe$_{2}$  monolayer in the FM ground state at tensile strains of (a) 2\%, (b) 6\%, (c) 10\%; PDOS of Cr 3$d$ and Te 5$p$ orbitals for the CrTe$_{2}$  monolayer in the FM ground state at tensile strains of (d) 0\%, (e) 6\%, (f) 10\% . The positive and negative values represent the DOS of up and down spins, respectively. The Fermi level is set to be 0 eV.}
\end{figure}

To further investigate the origin of the magnetic evolution under the applied strain, we plot in Fig. 6 the partial DOS (PDOS) of Cr $3d$ and Se $4p$ (Te $5p$) electrons for CrSe$_{2}$ (CrTe$_{2}$) monolayers at different tensile strains, at which the systems are in the FM states. It can be seen that the Cr $3d$ states are almost totally spin-polarized and become the main contribution of the magnetism, so the magnetic moments of Cr atoms are much larger than those of Se/Te atoms, as shown in Fig. 4(c). The hybridization between Cr $3d$ and Se $4p$ (Te $5p$) orbitals is strong for the spin-up states, indicating the covalent character of the Cr-Se or Cr-Te bonds. For the CrSe$_{2}$ monolayer, both the Cr $3d$ and Se $4p$ orbitals become more and more localized near the Fermi level when the strain is increased [from Fig. 6(a) to (c)]. At the strain of 10\%, the peaks of Cr $3d$ and Se $4p$ orbitals near the Fermi level appear at around $-$0.73 eV and 0.17 eV, respectively, and the hybridization of the two orbitals weakens, leading to more unpaired electrons near Cr and Se atoms. The more and more localized PDOS indicates the charge redistribution and the increased magnetic moments on both Cr and Se atoms, as shown in Fig. 4(c). It is noted that at the strain of 10\%, both the Cr $3d$ and Se $4p$ states become almost totally spin polarized and the CrSe$_{2}$ monolayer is nearly half metallic. We further increases the strain to 16\% and find that in the strain range of 12\% to 16\%, the system retains half-metallic. For the CrTe$_{2}$ monolayer, however, the spin polarization of Te $5p$ state becomes largest at the strain range of 4\% to 6\% and weakens when further increasing the strain, which is consistent with the result that the magnetic moment of Te atom is largest at this strain range. Since both the Cr $3d$ and Te $5p$ states cannot be totally spin-polarized, the CrTe$_{2}$ monolayer remains metallic for the strain range investigated.

\begin{figure}
\includegraphics[width=1.0\columnwidth]{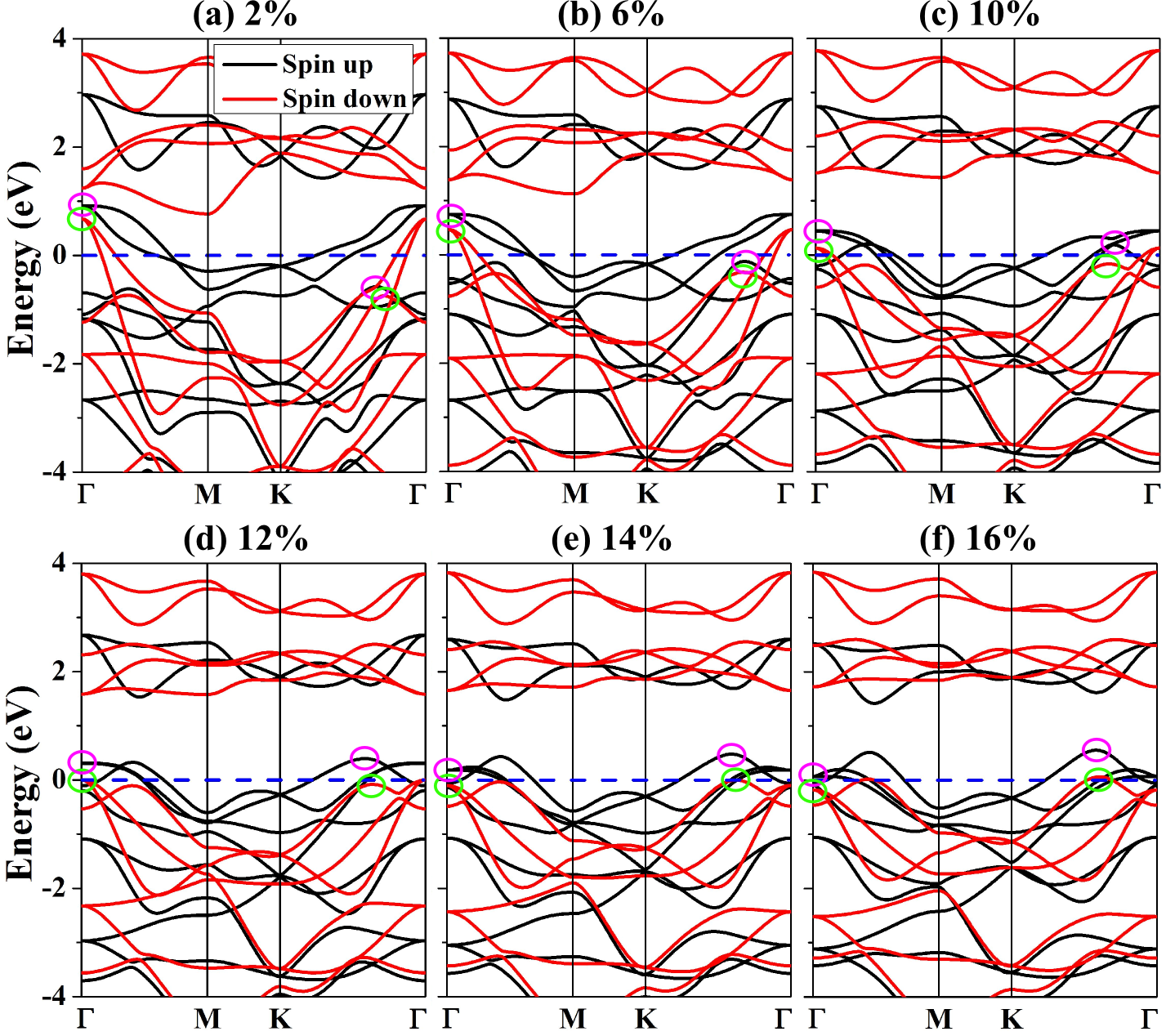}\caption{\label{fig7}Band structures of the CrSe$_{2}$  monolayer in the FM ground states at tensile strains of (a) 2\%, (b) 6\%, (c) 10\%, (d) 12\%, (e) 14\%, and (f) 16\%. The Fermi level is set to be 0 eV.}
\end{figure}

Two dimensional half-metallic ferromagnets are highly desirable for the future applications in high performance spintronic devices. To see clearly how the half-metallic property takes place in the CrSe$_{2}$ monolayer, we plot in Fig. 7 the band structures under different tensile strains, in which the system is in the FM ground states. For the spin-up energy band (drawn in black), the band extremum located at the $\Gamma$ point (circled by the purple line) is gradually lowered as the strain is increased, while the band extremum along the $K$-$\Gamma$ direction (also circled by the purple line) is elevated gradually. At the strain of 10\%, when the first band extremum has not been lowered below the $E_F$, the second band extremum has already moved up above the $E_F$ and keeps moving up as increasing the strain, so the spin-up state keeps metallic for the strain range investigated. For the spin-down energy band (drawn in red), however, the band extremum at the $\Gamma$ point (circled by the green line) is lowered much faster than that of the spin-up state when the strain is increased. At the strain of 12\%, the band extremum at the $\Gamma$ point moves below the $E_F$, with the band extremum along the $K$-$\Gamma$ direction also residing below the $E_F$, so the spin-down state tends to be semiconducting with a sizeable band gap of 1.58 eV. When the strain is further increased, the second extremum moves up very slowly and keeps below the $E_F$, so the CrSe$_{2}$ monolayer keeps half-metallic in the strain range of 12\% to 16\%.

\section{Conclusion}
In conclusion, we have investigated the magnetic properties of CrSe$_{2}$/CrTe$_{2}$ monolayers under the in-plane biaxial tensile and compressive strains. When no strain is applied, the CrSe$_{2}$ and CrTe$_{2}$ monolayers are found to be AFM and FM, respectively, consistent with their corresponding bulk counterparts. For the CrSe$_{2}$ monolayer, when the tensile strain reaches 2\%, the system tends to be FM. Furthermore, the FM state is stabilized and the Curie temperature $T_{\mbox{c}}$ as well as the magnetic moments increase with the increase of the tensile strain. A relatively small strain of 4\% can enhance the $T_{\mbox{c}}$ to be 363.1 K, above the room temperature. Moreover, the CrSe$_{2}$ monolayer changes from a metal to a half-metal when the strain is larger than 10\%. For the CrTe$_{2}$ monolayer, however, the critical strain at which the switch between FM and AFM states occurs is compressive, of $-1\%$. When the tensile strain is larger than 2\%, the $T_{\mbox{c}}$ of the CrTe$_{2}$ monolayer can be higher than the room temperature. The strain-tunable magnetic properties of CrSe$_{2}$ (CrTe$_{2}$) monolayers can be understood by the competition between the AFM Cr-Cr direct exchange interaction and the FM Cr-Se(Te)-Cr superexchange interaction. Our results indicate that the strain is an effective way to tune the magnetic and electronic properties of CrSe$_{2}$/CrTe$_{2}$ monolayers. The two monolayers are potential candidates for the future nanoelectronic applications, which deserve further study in the experiment.

\section{Acknowledgement}

This work was supported by the National Key Basic Research under Contract No. 2011CBA00111, the National Natural Science Foundation of China under Contract Nos. 11274311 and 11404340, the Joint Funds of the National Natural Science Foundation of China and the Chinese Academy of Sciences' Large-scale Scientific Facility (Grant No. U1232139), the Anhui Provincial Natural Science Foundation under Contract No. 1408085MA11, the China Postdoctoral Science Foundation (Grant No. 2014M550352) and the Special Financial Grant from the China Postdoctoral Science Foundation (Grant No. 2015T80670). The calculation was partially performed at the Center for Computational Science, CASHIPS.

\end{document}